\documentclass[aps,prb,twocolumn]{revtex4}
\begin{document}
%\newlength\FigWidth
%\FigWidth 3.25 true in
%\FigWidth 5 true in
%\draft
\title{Theory of a Higher Order Phase Transition: Superconducting 
Transition in BKBO}
\author{Pradeep Kumar}
\affiliation{Department of Physics, PO Box 118440, University of Florida, Gainesville, FL 32611-8440}
\date{\today}
\begin{abstract}
We describe here the properties expected of a higher (with emphasis on the order fourth) order phase transition.  The order is identified in the sense first noted by Ehrenfest, namely in terms of the temperature dependence of the ordered state free energy near the phase boundary.  We have derived an equation for the phase boundary in terms of the discontinuities in thermodynamic observables, developed a Ginzburg-Landau free energy and studied the thermodynamic and magnetic properties.  We also discuss the current status of experiments on $Ba_{0.6}K_{0.4}BiO_3$ and other $BiO_3$ based superconductors, the expectations for parameters and examine alternative explanations of the experimental results.
\end{abstract}
\maketitle

\section{Introduction}
There has been a proposal\cite{kumar1, hall} recently that the transition to superconductivity in the cubic, non-copper based,  moderate $T_c$ (32K) material $Ba_{0.6}K_{0.4}BiO_3 (BKBO)$ is of order four in the sense proposed by Ehrenfest\cite{ehren}. The free energy and its first three derivatives are continuous across the phase transition. The fourth derivatives with respect to temperature (or, alternatively, with respect to some mechanical variable such as pressure or magnetic field) are discontinuous.  This is somewhat unexpected. After nearly a century of an extensive study of phase transitions, where the transitions were either discontinuous (in the Landau sense of a discontinuous change in the order parameter and/or entropy, the latter in the sense Ehrenfest envisioned) and therefore first order, or continuous where the transition was believed to be II order, a consensus seems to have emerged that all departures from a strictly II order behavior can always be explained in terms of thermodynamic fluctuations. 

To recall, the Ehrenfest definition of the order of a phase transition can be summarized in the thermodynamic free energy $F(B,T)$. Here $B$ is the magnetic field and $T$ is the temperature. Thus near the phase boundary (for a superconductor) the free energy satisfies:

\begin{equation}
\label{eq1}
 -F(B,T) \approx [1 - T/T_{c}]^{p - \mu} \approx [1 - B/B_{c2}]^{p -\zeta} 
\end{equation}

Here p is an integer, the order of the transition and $\mu$ and $\zeta$ are small corrections. The scaling field $B_{c2} \approx a^{x_2}, (a = a_o(1-T/T_c))$ defines the exponent $x_2$. Thus all derivatives of this free energy, of order less than p, will be continuous, the $p^{th}$ derivative will be discontinuous if $\mu$ and $\zeta$ are zero. Otherwise they will be divergent with exponents $\mu$ for a thermal derivative and $x_2\zeta$ for a field derivative. For a second order phase transition, p = 2. The lowest thermal derivative, which may be singular, is the specific heat (or the magnetic susceptibility). Apart from conventional superconductors and fermionic superfluids where mean field theory is valid and $\mu$ and $\zeta$ are both nearly zero, most transitions are accompanied by a small but finite $\mu$.

That $\mu$ is small, is on the one hand a validation of the Ehrenfest classification. On the other hand, it facilitates formalism, such as the renormalization group, to calculate the numerical value of $\mu$ and $\zeta$, depending on symmetry properties alone.  Disorder brings about an uncertainty to this clean picture.  Imagine for example a system with a distribution of $T_c$'s caused by spatial inhomogenieties.  The goal here is to establish that a broad transition may be more than just a dirt effect.  The transition may also be higher order in that the exponent p above is larger than 2.  It might be tempting to use the known formalism for a second order transition to calculate larger (in magnitude, since they are negative) values of a $\mu $.  However, since the mean field theory for a higher order transition is different, the calculations should appropriately be carried out within the framework of a higher order transition.

In a second order phase transition, the slope of the phase boundary in the B-T plane is related to the discontinuity in specific heat ($\Delta C$) and in the magnetic susceptibility ($\Delta \chi$),

\begin{equation}
\label{eq2}
\Delta C = T_{c} \Delta \chi {\left[{{\frac{{dB_{c2}} }{{dT}}}} 
\right]}^{2}
\end{equation}
If $\Delta C$ is zero then either the phase boundary has a zero slope or $\Delta \chi$ is also zero and the phase transition is of higher order.  There are materials where $\Delta C = 0$ at the superconducting transition.

The class of cubic perovskite materials, the bismuthates, seem to be the promising candidates for the effects discussed here.  Following the discovery\cite{sleight} of superconductivity at 11K in $BaPb_xBi_{1-x}O_3$(BPBO), there has been a continuing interest in this class of materials. While BKBO keeps the record\cite{cava} for the highest $T_c$ (32K) and the Rubidium-doped samples $Ba_{1-x}Rb_x BiO_3$ (BRBO, $T_c$=25K)\cite{tomino, kuentzler} have $T_c$'s which are  slightly smaller, most of the materials in this series have a transition temperature nearer to 10K.   They are all missing the specific heat anomalies.

The new materials in this class are $K$ and $Rb$ doped $SrBiO_3$ with $T_c = 12-13K$\cite{kazakov} and $K_{1-x}Bi_{1+x}O_3$ with a $T_c$ = 10K \cite{khasanova}. The straggler in this group is $BPSO$ $(Ba(Pb,Sb)O_3)$\cite{batlogg} with the highest $T_c$ of 3.5K, but otherwise with similar properties as the corresponding ones in the other members of the group. These materials await further work such as a careful measurement of specific heat and magnetic properties as a function of magnetic field and pressure.  

The natural question here is what happens if the two discontinuities are simply very small.  This is similar to what happens in a second order phase transition which turns weakly first order.  The entropy should have a kink, instead it undergoes a small jump.  Here, too, the system approaches with all the accoutrement (for example the temperature and field dependence of the free energy as noted above in Eq.(\ref{eq1})) of a higher order transition, but then turns weakly second order.  The suggestion below is that the order of the transition should be identified from the temperature dependence of the free energy as one approaches the transition.

This paper is a follow-up to an earlier brief letter\cite{kumar1} on the report of and an interpretation of the anomalous superconducting transition in BKBO. It is also a theoretical complement to a detailed report of the experiments\cite{hall}. The conclusions of ref.(\onlinecite{kumar1}) are based on three independent results in the magnetic field and temperature dependent magnetization of BKBO. They are (recall that $a$ here refers to the reduced temperature $(1-T/T_c))$:

1. The thermodynamic critical field has the temperature dependence (all fields are in Tesla) $B_o(T)= 0.51 a^{1.81}$.

2. The lower critical field has the temperature dependence with a cross-over.  Near $T_c$, it is $B_{c1}(T) = 0.096 a^{3.03}$. At low T, the temperature dependence is linear in $a$.

3. The upper critical field $B_{c2}(T) = 19.7 a^{1.21}$. The Ginzburg-Landau parameter $\kappa = \sqrt{B_{c2}/B_{c1}} = 14.4 a^{0.91}$. 

These results, as we discuss below, challenge the view that the phase transition to superconducting state is of second order.  They instead lead to an internally consistent description of a fourth order phase transition. Result (1), in the context of Eq. (\ref{eq1}) shows that p = 4 and $\mu = 0.4$. Result (2) is a consistent with the field theory developed for a fourth order phase transition.  In general, we can show that for a $p^{th}$ order transition, London penetration depth (and therefore $H_{c1}$) satisfies $\lambda^{-2} \approx a^{(p-1)}$.

We can also show that within mean field theory, the superconducting coherence length satisfies $\xi^{-2} \approx a$.  Thus the parameter $\kappa = \lambda/\xi$ is constant only for a II order phase transition.  In a transition of order higher than 2, $\kappa$ diverges at $T_c$.  Physically, one might view this as a separation between the magnitudes of flux penetration and order parameter variation.  Near $T_c$, flux expulsion gets weaker for a higher order transtion.  Althought as we see below in Sec. (IV), the estimates derived for low temperatures  correspond to $\kappa \approx 14$, a large value.

The paper is organized as follows. In sec. II, we discuss the thermodynamic foundations of a fourth order phase transition. Sec. III contains discussion of a model field theory for a higher order phase transition. Sec. IV follows up with a discussion of the gradient terms in the field theory under the heading: magnetic field effects. Sec. V is a discussion of the reported results and the existing estimates of the specific heat discontuity.  While they seem to agree, there are reasons to believe that the agreement might be conincidental. Sec. VI concludes with a summary of principal results and a discussion of assumptions. 

\section{Thermodynamics of a Higher Order Transition}

In the Ehrenfest definition of the order of a phase transition, $p^{th}$ order phase transition is characterized by a discontinuity in the $p^{th}$ derivative of the free energy. Consider the free energy as a function of the temperature $T$ and a mechanical variable, say the magnetic field $B$. The discontinuities also determine the phase boundary in the $B-T$ plane. When p=1, the discontinuous derivatives are the entropy $S$ and the magnetization $M$.  We have the Clausius-Clapeyron equation: dB/dT = - $\Delta S/\Delta M$.  If the entropy and the magnetization are continuous across the phase boundary, the transition may be of higher order.  For p=2, $[dB/dT]^2 = \Delta C/T_c\Delta \chi$.

It is possible to generalize these results.  For a third order phase transition, the corresponding relation becomes 
\begin{equation}
\label{eq4}
\left[{{dB}\over {dT}} \right]^3 = - {{\Delta 
{{\partial C}/{\partial T}}} \over
{T_{c} \Delta {{\partial \chi}/{\partial B}}}}
\end{equation}
To derive this relation, note that in the vicinity of the phase boundary separating phases 1 and 2; l'Hopital's rule will necessarily change Eq. (\ref{eq2}) in to 
\begin{equation}
\label{eq5}
\left[{{dB}\over {dT}} \right]^2 = {{\Delta 
{{\partial C}/{\partial T}}} \over
{T_{c} \Delta {{\partial \chi}/{\partial T}}}}
\end{equation}
Since the susceptibility $\chi$ is now continuous across the transition, it follows that;
\begin{equation}
\label{eq6}
{{dB}\over {dT}} = - {{\Delta 
{{\partial \chi}/{\partial T}}} \over
{\Delta {{\partial \chi}/{\partial B}}}}
\end{equation}
Multiplying Eqs.(\ref{eq5}) and (\ref{eq6}), we recover Eq.(\ref{eq4}).

Similarly, the expression for a fourth order phase transition becomes:

\begin{equation}
\label{eq7}
\left[{{dB}\over {dT}} \right]^4 = {{\Delta 
{{\partial^2 C}/{\partial T^2}}} \over
{T_{c} \Delta {{\partial^2 \chi}/{\partial B^2}}}}
\end{equation}

A generalization to an arbitrary order phase transition is nearly self-evident here. The proof is by induction, just as we derive Eq.(\ref{eq7}) from (\ref{eq4}) or Eq.(\ref{eq4}) from Eq.(\ref{eq2}).  

\begin{equation}
\label{eq71}
\left[{{dB}\over {dT}} \right]^p = (-1)^p{\Delta 
{{\partial^{p-2} C}/{\partial T^{p-2}}} 
\over
{T_{c} \Delta {{\partial^{p-2} \chi}/{\partial B^{p-2}}}}}
\end{equation}

The expression is not unique, although certain of its aspects are notable such that (a) the slope is raised to the power of the order of the transition and (b) that the numerator consists of the discontinuity in the thermal derivative of the free energy while the denominator contains the derivative with respect to the mechanical variable, here the magnetic field. Of course, there are other expressions involving mixed thermal and mechanical derivatives which have lower powers of the slope dB/dT. There are also relations between 
mixed derivatives corresponding to the well known Maxwell relations. The expressions above seem to be the most compact and symmetric. 

It is possible to write the corresponding expressions for the case when the mechanical variable is pressure. For a second order phase transition, the density remains the same across the transition but the compressibility, the second derivative of the free energy as a function of pressure is discontinuous. Since the longitudinal sound velocity directly involves the compressibility, if the transition temperature depends on the pressure, the sound velocity changes at the transition. In a fourth order phase transition, the compressibility is continuous across the transition as well as its first derivative, but its second derivative, with respect to both the temperature or the pressure is discontinuous and should be observable in the temperature/pressure dependence of the sound velocity..

In the above, the implicit assumption has been that $\mu$ and $\zeta$ are zero.  If they are nonzero, then the discontinuities in free energy derivatives turn into weak logarithmic divergences. One can then take the exponents for these divergences and derive scaling laws between them.  Assuming that the magnetic field and temperature dependences in a $p^{th}$ order phase transition are described by Eq. (\ref{eq1}), Kumar and Saxena\cite{kumar2} have derived the scaling laws. Defining the exponents as:

$${{\partial^{p-2} C}/{\partial T^{p-2}}} = a^{-\mu};\quad {{\partial^{p-2} \chi}/{\partial B^{p-2}}} = a^{-\kappa}$$
$$ M = -\frac{\partial F}{\partial B} = a^{\beta}; \quad m(T=T_c)=B^{1/\delta}$$ 

\begin{equation}
\label{eq72}
(p-1)\mu +p\beta +\kappa = p(p-1)
\end{equation}
\begin{equation}
\label{eq73}
\kappa = \beta[(p-1)\delta -1]
\end{equation}
\begin{equation}
\label{eq74}
\mu=p-\beta[\delta+1]
\end{equation}
\begin{equation}
\label{eq75}
(\delta+1)\kappa+[(p-1)\delta-1](\mu-p)=0
\end{equation}
This brings us to the end of the section on thermodynamics.  The next section develops possible alternatives for a field theory, underlying the thermodynamic properties. 

\section{A Model Field Theory}

We need to construct a free energy $F$\{$\psi $\}, a function of the complex order parameter $\psi  =\Delta e ^{i\phi}$ such that when $F$ is minimized with respect to the order parameter $\psi $, its minimum is the thermodynamic free energy.  $F$ is also a measure of the fluctuations, i.e. those configurations of $\psi (r)$ which are near the equilibriunm configuration $\psi_o (r) $. In the spirit of Ginzburg-Landau formalism, $F$ could be written as an expansion in the powers (even powers, since $F$ must be gauge invariant; a change in $\phi$ to $\phi+c$, where $c$ is a constant, should leave $F$ invariant) of $\psi $. However such an expansion necessarily leads us to a second order phase transition. 

To consider a different temperature dependnece of the free energy, suppose we write the leading term in $F \approx a \psi ^{2t}$ (recall here that $a = a_o(1-T/T_c)$ and $t$ is defined here as an exponent), then to be consistent with Eq. (\ref{eq1}), we need to make sure that $|\psi| ^{2t} = a^{(p-1)}$. This can be accomplished in seemingly two different ways. First consider the choice, $\psi ^2  \approx  a$, then t = (p-1). The free energy for a fourth order phase transition, in this approach can be written as

\begin{equation}
\label{eq8}
F\{\psi = \Delta e^{i\phi}\} = - a\Delta ^{6} + b\Delta ^{8} + F_{G} 
\end{equation}
where $F_G$ represents the terms of various powers in $|\nabla \psi |^2$. The second term with the power 8 ensures the first of the conditions, that $|\psi| ^2  = a$.  

The second possibility is provided by the choice t = 1, leading to a free energy (to avoid confusion, here we use $n$ for the order parameter),

\begin{equation}
\label{eq9}
F\{n\} = - an^{2} + bn^{8 / 3} + F_{G} \{n\}
\end{equation}

Eqs. (\ref{eq8}) and (\ref{eq9}) are related via the transformation $n = \psi^3 $.  An ambiguity here is the choice of the cube, for example, the transformation could well be $n = |\psi|^2\psi$, then the phases for $\psi$ and $n$ are identical.  We have no way of deciding on this choice.  The reader is left to speculate\cite{note3} about the possibility of a fractional flux quantum.

So far we have not considered the gradient terms.  In Eq. (\ref{eq9}), it is clear that it must be c$| \nabla n|^2$. Transformed back into Eq. (\ref{eq8}), the gradient term becomes $| \nabla {\psi} ^ 3|^2$. Even in Eq.(\ref{eq8}) it can be argued that the gradient term must be as described here.  If it were not and say the gradient term were simply $|\nabla \psi|^2$, the temperature dependence of this term would be $a^2$.  In the presence of a magnetic field the transition would switch to a second order.  Assuming that $\nabla^2 \approx a$, the gradient terms consistent with a fourth order transition are,

\begin{equation}
\label{eq91}
c_1|\nabla \psi^3|^2+c_2|\nabla^2 \psi^2|^2 +c_3|\nabla^3\psi|^2
\end{equation}

The additional terms are indicative of a nonlinear Meissner effect and will be omitted from any further discussion of stability effects. 

We now have a free energy which can also be written as

\begin{equation}
\label{eq10}
F\{\psi (r),T\} = \int {dV\psi ^{4}{\left[ { - a\psi ^{2} + b\psi ^{4} + 
c\vert \nabla \psi \vert ^{2}} \right]}} 
\end{equation}

This looks like the usual free energy, multiplied by an external weight factor proportional to $\psi^4 $.  In the BCS\cite{schrieffer, tinkham} derivation of the Ginzburg-Landau free energy, the prefactor is the density of states, which determines the energy scale of the phenomena.  Here the extra factor may be seen as arising from a density of states which is proportional to $\psi^4 $.  Thus the effective density of states in the normal state vanishes. It is non-zero only in the condensed state where it has a given temperature dependence related to the order of the phase transition. In effect we have a transition from an insulator to a superconductor.

Szabo et al\cite{szabo}have measured the temperature dependence of the energy gap by point-contact tunneling with a silver electrode. They find a BCS like temperature dependence.  This would be consistent with the description above if  $\Delta $ were to be identified as the energy gap (or more precisely, it rules out the possibility of $n$ being identified as an energy gap).

Let us also note that the free energy for a third order phase transition is given by (corresponding to Eqs. (\ref{eq8}) and (\ref{eq9}), 

\begin{equation}
\label{eq11}
\begin{array}{l}
 F_{3} = - a\Delta ^{4} + b\Delta ^{6} + c|\nabla \psi ^2|^2 
\\ 
 = - an^{2} + bn^{3} + c|\nabla n|^{2} \\ 
 \end{array}
\end{equation}
In the above equations, the relation is n = $\psi ^2$.

Finally let us note that the interaction term with the power 8/3 in Eq. (\ref{eq9}) may well arise from an averaging carried out over some other field. In the spirit of a Hubbard-Stratonovich transformation, there may be auxiliary fields which interact with the superconducting order parameter and are ultimately responsible for the unusual order of the transition. Looking at Eq.(\ref{eq8}), one might get the impression that it is necessary to find some physical reason for the absence of terms quadratic and quartic in $\psi $.  The answer is to some extent, provided by Eqs. (\ref{eq9},\ref{eq10}).  The primary message here is to point out that there are many different ways to see the mathematical structure necessary for a higher order phase transition.

In a real physical system, there are often nearby instabilities. For example in $BaBiO_3$, there are several charge-density-wave (CDW)\cite{taraphder, rosenfeld} transitions (see for example Ref.(\onlinecite{kuentzler})). The corresponding transition temperatures decrease with K concentration, x and the highest of them vanishes near x=0.4 . This value of x coincides with the highest superconducting $T_c$. A microscopic theory may well involve a subtle interplay between the CDW and superconducting degrees of freedom.

To conclude this section, note the results arising from free energy minimization of Eq.(\ref{eq9}).

\begin{equation}
\label{eq111}
 \Delta _o ^{2} = {{3a}\over {4b}};\quad for\quad T < T_c  
\end{equation}

The thermodynamic free energy (the free energy at the minimum) $F_o(T)$ is given by;

\begin{equation}
\label{eq112}
 F_{o} (T) = - {{27}\over {250}} {{{a^4}\over {b^3}}} + F_N (T)
\end{equation} 
 $$\Delta C = C(T) - C_{N} (T) = - T{{\partial ^2}\over {\partial T^2}}(F_{o} - F_{N} )\\$$ 
\begin{equation} 
\label{113}
\Delta C = {{81}\over{64}}{{a_o ^4}\over {T_{c} b^3}}(1 - T / T_c)^{2} \quad for\quad T<T_c \\ 
\end{equation}

The specific heat approaches $T_c$ with a power law. Here for a 
fourth order transition, the power is quadratic. There is a discontinuity in $\partial^2 C/\partial T^2$ at $T_c$;

\[
\Delta {\frac{{\partial ^{2}C}}{{\partial T^{2}}}} = 
{\frac{{81}}{{32}}}{\frac{{a_{o} ^{4}}}{{T_{c} ^{3}b^{3}}}}
\]

\section{Magnetic Field Effects}

This section addresses the consequences of the gradient terms in the free energy in Eqs.(\ref{eq8},\ref{eq9}). In a charged superconductor, the effect of magnetic field appears via the gauge transformation, $\nabla  \rightarrow \nabla + 2\pi i A/\varphi_o $, where $A$ is the vector potential and $\varphi _o = h/2e$ is the superconducting flux quantum. The free energy Eq. (\ref{eq10}) is also supplemented by the field energy density term, $B^2/2\mu$.  Focussing on issues specific to this section, note that the London equation is the Euler-Lagrange equation for the vector potential $A(r)$. Following Eq. (\ref{eq8}),

\begin{equation}
\label{eq12}
0 = {\frac{{\partial F}}{{\partial A}}} \Rightarrow - {\frac{{1}}{{2\mu _{o} 
}}}\nabla ^{2}A + c({\frac{{2\pi} }{{\varphi _{o}} }})^{2}\Delta ^{6}A = 0
\end{equation}

The flux is expelled whenever $\Delta \neq 0$. There is a corresponding 
penetration length, $\lambda$ such that (recall that $a$ is the reduced 
temperature, $a \approx (1-T/T_c)$)

\begin{equation}
\label{eq13}
\lambda ^{ - 2}(T) = 2\mu _{o} c\left( {2\pi / \varphi _{o}}  
\right)^{2}\Delta ^{6} \cong a^{3}
\end{equation}

This is a mean field result for a fourth order phase transition.  For a general order $p$, of the phase transition, the temperature dependence of $\lambda ^{-2} \approx a^{(p-1)}$.  This is quite distinct from the fluctuation contributions to the exponent $x_1$ (defined as $\lambda^{-2} \approx a^{x_1}$).  For a second order transition, the classical value of $x_1$ = 1 but the fluctuations tend to make it smaller. For example, for a 3d XY model, the superfluid density exponent $x_1 \approx 2/3$, measured\cite{kamal} in several high $T_c$ superconductors.  For a $p^{th}$ order transition, fluctuations will probably reduce it to $ \le p-1$ but not by much and the connection to the classical value and the corresponding order, as well as the contrast with 3d XY model should remain clear.

The lower critical field\cite{tinkham} in a superconductor $B_{c1}(T) \approx \varphi_{o}/\lambda ^2 \approx a^{(p -1)}$. This should be a clear observable.  Indeed the expression for $B_{c1}(T)$ contains corrections logarithmic in $\kappa = \lambda/ \xi $ where $\xi $ is the temperature dependent coherence length. In a second order phase transition, both $\lambda $ and $\xi $ have the same temperature dependence making $\kappa $ a constant. In a higher order transition, $\kappa$ is temperature dependent and thus the temperature dependent exponent for $B_{c1}(T)$ has logarithmic corrections; even at the mean-field level, it may be slightly smaller than $(p-1)$.

The divergence of $\kappa$ indicates a novel physical effect. As the condensation occurs, the flux expulsion takes place more gradually than the development of order parameter stiffness.  There is no reason that they should happen together (as they do when the transition is of second order).  Indeed it could be argued that the curious behavior is the constancy of $\kappa $. The range of values of $\kappa$ in BKBO is such that the system is always a superconductor of type II, even at low temperatures. This may also explain why the measured values of $\kappa $ vary so much between different experiments. They have been measured at different temperatures.

Let us note that for a third order transition, the temperature dependence of  $\lambda ^{-2} \approx a^2$.  If the transition in Bi-2212 is of third order,\cite{junod} then the penetration depth will correspond to a different power law. The exponent may be smaller than 2 due to the fluctuations, but it would be considerably larger than one, the number characteristic of a second order phase transition. For a gas of bosons, Schafroth\cite{schafroth} has derived a temperature dependence for $\lambda ^{-2}  \quad  \approx a^2\ln a$. There are conflicting reports of two experiments, one noted\cite{jacobs} the temperature dependence as quadratic in the reduced temperature, the other noted\cite{waldram} its absence.

Based on the experimental results\cite{hall} for BKBO, which include the lower critical field, $B_{c1}(T) = 0.0955 a^{3.03}$ Tesla, the upper critical field $B_{c2} = 19.7a^{1.21}$ Tesla and the thermodynamic critical field $B_o = 0.51 a^{1.81}$ Tesla, we can derive the characteristic limiting length scales at T = 0. 

\begin{equation}
\label{eq15}
\begin{array}{l}
 \lambda(0) = \sqrt {{\frac{{\varphi _{o} / 2\pi} }{{B_{c1} (0)}}}} = 830\AA \\ 
 \xi(0) = \sqrt {{\frac{{\varphi _{o} / 2\pi} }{{B_{c2} (0)}}}} = 58\AA \\ 
 \end{array}
\end{equation}

Finally, in a second order phase transition superconductor, there is an identity $B_o^2  = B_{c1} B_{c2}$ . This identity is not satisfied here. The geometric mean of the two critical fields (in Tesla) is $1.37 a^{2.12}$ while the thermodynamic field, as noted above is $0.51 a^{1.81}$. There is no clear reason to draw any conclusions from this observation.  It may be that the identity is simply not expected to be valid in general.

Finally, let us note that the Euler-Lagrange equation for the order parameter is simply derived,

\begin{equation}
\label{eq14}
0 = -6a\psi +8b\psi^3 - 4{|\nabla\psi|^2 \over \psi} - 2\nabla^2 \psi
\end{equation}
We will defer a detailed study of the consequences of this equation to until later.

\section{Discussion of Experiments}

In retrospect, it is interesting to note that soon after the original discovery\cite{sleight} of superconductivity in BPBO by Sleight et al, Methfessel et al\cite{meth} published a paper with the title "Why is there no bulk specific heat anomaly at the superconducting transition temperature of BPBO". The authors  first wondered about the nature of superconductivity in BPBO, whether it was some sort of impurity phase since magnetic measurements of superconducting transition have often turned up to be misleading. But they also noted that "if it is a bulk effect, then we may have to invoke a new, as yet unidentified, mechanism for its cause". Soon after the discovery of superconductivity in BKBO by Cava et al\cite{cava}, the early specific heat measurements\cite{hundley, stupp} were already reporting the missing specific heat anomaly. There was one report\cite{graebner} of a small specific heat discontinuity but the two other measurements were unable to verify it. 

It is possible to estimate the size of the expected discontinuity in specific heat. When this question is put before the cognoscenti, the answer one is more likely to get is that the specific heat discontinuity in BKBO is small, of the order of a few mJ/moleK. This estimate, as we discuss below, is based on questionable assumptions.  Most of the reports of measurements of this magnitude are less than convincing. 

We show below that the estimate is based on arguable evidence. For instance, several authors\cite{kwok} have used formulae based on thermodynamic identities for a second order phase transition, in one case, a version of the familiar Ehrenfest expression, Eq. (\ref{eq2}) where the identity $\Delta \chi = [8 \pi \kappa^{2}]^{-1}$ converts this relation into an expression for $\Delta C$ in terms of experimentally known quantities $\kappa$ and the slope of the $B_{c2}$ curve near $T_c$.  Kwok et al\cite{kwok} used this expression to estimate $\Delta C/T_c \approx 3.75$ mJ/mole/K$^2$. In the measurements of Hall et al\cite{kumar1,hall}, $\kappa$ apparently diverges near $T_c$ as $\kappa \approx  (1-t)^{-1}$. If we use this temperature dependence, we get $\Delta C$ = 0. 

Similarly, based on a more fundamental expression, 

\begin{equation}
\label{eq3}
\Delta C = T_{c} {\frac{{\partial ^{2}}}{{\partial T^2}}}\left( {{\frac{{B_{o} 
^{2}}}{{2\mu _{o}} }}} \right)
\end{equation}
\noindent
where $B_o$ is the thermodynamic critical field. Batlogg et al\cite{batlogg} estimated $\Delta C/T_c \approx 2.1\pm 0.4$ mJ/mole/K$^2$. Once again, if we use the experimentally\cite{kumar1,hall} measured temperature dependence of $B_o \propto (1-T/T_c)^2$, we get $\Delta C = 0$.  One begins to get the impression that the small value of $\Delta C$ may well be an imperfect estimate of zero.

On the other side of this story, there is an estimate which yields an expected $\Delta C$ at least three orders of magnitude larger.  According to Hundley et al\cite{hundley, stupp}, the normal state specific heat in BKBO can be fit to an electronic contribution of $C/T = \gamma = 0.15$  J/mole K$^2$. The data, when plotted as $C/T vs. T^2$, show two straight lines with a crossover around 16K. On the low temperature side, the T=0 intercept is zero. However on the high temperature side ($>16K$, transcending $T_c$), the intercept has the value quoted above. If we take the view that the low temperature side is the property\cite{note2}  of the superconducting state, then the effective $\gamma$ derived from the high temperature side must be the large value. Since $\Delta C \approx \gamma T_c$, we expect a {\it{discontinuity}} at the superconducting transition of the order of 4-6 J/moleK.

 In a recent letter, Woodfield et al\cite{woodfield} have questioned all of the conclusions of ref. (\onlinecite{kumar1}) on the grounds that they (Woodfield et al) have measured a $\gamma = 0.9  mJ/moleK^2$ and $\Delta C/T_c = 2.1 mJ/moleK^2$, a result in excellent agreement with the estimates of Batlogg et al. But, as the discussion above suggests, there may be a different point of view for these small estimates for $\Delta C$.  In any case, a non-vanishing discontinuity in a higher derivative of $\Delta C$ still leads to a structure below $T_c$.  For a fourth order phase transition, the discontinuity in second thermal derivative of the specific heat means that the specific heat rises below $T_c$, although quadratically in the reduced temperature $a$.  It eventually vanishes at $T=0$.  That means that there must be a specific heat maximum below the transition temperature.  It is just that the maximum (or some measure related to it, such as the temperature half way to it) should not be interpreted as an estimate of $T_c$.  Nor is the height of the maximum to be interpreted as an estimate of the discontinuity. Indeed, whereas in a second order phase transition, the size of the discontinuity is a condensate property, the corresponding information in a fourth order transition lies in the coefficient of the $a^2$ term in the specific heat.

There is a report from the Grenoble group\cite{blanchard} of measurements of specific heat and the magnetization in BKBO.  This report emphasizes the presence of a specific heat maximum below $T_c$, the latter derived from tunneling measurements associated with a vanishing energy gap.  Near $T_c$, the specific heat appears to be quadratic in $a$.  Moreover, they have obtained a measure of $\lambda^{-2}$ from the magnetization which seems to fit $a^{1.5}$.  If the transition is second order, the mean field exponent is one.  Fluctuations, within a 3d XY model yield\cite{kamal} smaller exponent (of order 2/3).  This result is also in sharp contrast to the results of Hall\cite{kumar1,hall}, where the exponent is closer to 3.  This discrepancy, amongst experiments, of a factor of 2 in the exponent, is unexpected and needs to be addressed. 

There are reports of energy gap measurements from tunneling experiments\cite{samuely, szabo} which satisfy the BCS characteristics. Thus the ratio $2\Delta /kT_c$ is in the range of $4-4.3$, only slightly larger than the BCS value (3.4) and possibly indicative of strong coupling effects. There is no reason to expect higher order transition effects in the temperature dependence of the order parameter (which is presumably also the energy gap). For example the temperature dependence of the energy gap measured by Szabo et al \cite{szabo} very well fits BCS expectations.  Eq.(\ref{eq10}, describes an order parameter which is conventional in its temperature depndence, but the free energy temperature dependence is not.
 
\section{Summary and Conclusions}

To recall, perhaps the more convenient definition of the order of a transition is the one based on Eq. (\ref{eq1}), namely it is the integer p, in the exponent (${p-\mu}$ in the temperature dependence of the free energy.   This leads to the following conclusions.

(1) The superconducting phase transition in BKBO has the characteristics of a IV order phase transition in the sense described by Ehrenfest.  This indentification is based on the temperature dependence of the themodynamc critical field, the temperature dependence of the lower critical field (or the London penetration length) and is consistent with the missing discontinuity in magnetic susceptibility and the specific heat.

(2) The equation for the phase boundary in a $p^{th}$-order phase transtion is described by Eq. (\ref{eq71}).  The corresponding relation for a $4^{th}$- order transition, Eq. (\ref{eq7}) relates the lowest order singular derivatives of the free energy. 

 (3) A model field theory, consisting of polynomials in the order parameter magnitude, is able to describe the anomalous temperature dependences in item (1) above.  In particular, the free energy leads to the anomalous temperature dependence of the thermodynamic as well as the lower critical field.  This free energy may arise in the vicinity of competing phase transitions 

The possibility of a higher order phase transition becomes viable if the specific heat discontiuity is absent.  In BKBO\cite{hundley, stupp}, as well as in BRBO\cite{kuentzler} (Rubidium instead of Potassium) and in BPBO (Pb replacing Bi)\cite{meth}, the specific heat discontinuity is anomalous, either missing or believed to be extremely small.  Several discussions in the literature advocate the belief that, in BKBO, $\Delta C$ is indeed expected to be quite small.  We have analyzed these estimates and find that the small estimates of   $\Delta C$ may well be imperfect estimates of zero.   When combined with the other indications (anomalous temperature dependence of the thermodynamic critical field and of the lower critical field as well as the missing discontinuity in the magnetic susceptibility) the evidence for an unusual transition is notable.  This evidence is strictly thermodynamic.  The line of reasoning is model-independent.

It is possible to understand this transition in terms of a model free energy whose expansion in powers of the order parameter begins with the $6^{th}$ order term.  If we, as is commonly done in textbooks, assume that
all phase transition must be described by a Ginzburg-Landau free energy with an expansion in all powers of the order parameter (including quadratic and quartic), then this assertion is equivalent to saying that higher order phase transitions do not exist.  That the thermodynamic evidence described here is flawed in some unknown but fundamental way and must be rejected.  If on the other hand, the experimental evidence for anomalous temperature dependence stands the test of further experiments, then we may have to abandon the "all powers" requirement for the free energy.   One possibility might be gleaned from Eq. (\ref{eq10}).  Here the free energy density is an "all powers" expansion, although the total free energy is an integral over all space with a "weight" function.

There remains the task of identifying the microscopic origins of the phenomena discussed here.   At this stage the experimental evidence does not provide clear insight into an answer to this question.  Certainly, future experiments will.  Hopefully the above discussion will stimulate more experiments in future.

\section{Acknowledgments}

I am grateful to many for discussions, encouragement,criticism  and support.  The list of names must include, R. G. Goodrich, Donavan Hall, A. J. Houghton, A. Migliori, W. Saslow, J. Sauls, A. B. Saxena, J. R. Schrieffer and H. Suhl.  The National Science Foundation has supported my work over the past five years in a number of ways.

%\newpage 
%{\bf References} 


\begin {thebibliography}{99}
\bibitem{note1} There have been earlier reports. The best known perhaps is the observation that Bose-Einstein condensation (in an ideal Bose gas), with a specific heat that has a kink at T$_c$, is a third order phase transition.  This is not in accord with the Ehrenfest classification where in the [P-T] plane, (a) the specific heat C and also the compressibility $\kappa$ must be continuous at T$_ c$ and (b) the temperature derivative of C and the volume derivative of 
$\kappa$ should be discontinuous for a third order phase transition. In BEC in an ideal Bose gas, the compressiblitiy is infinite and London has argued that the transition should really be viewed as a first order phase transition. Other reports of a higher order transition in Bi and Tl based high-T$_ c$ oxides are based largely on specific heat alone. Without the corresponding results for magnetization and susceptibility, it is premature to draw definitive conclusions. Although see the later discussion about Bi-2212 and its specific heat.
\end{thebibliography}
\end{document}